\documentclass[varenna]{cimento}

\setcopyright{CERN on behalf the ALICE Collaboration}
\usepackage{comment} 
\usepackage{amsmath,amssymb}
\usepackage{lineno}
\usepackage{graphicx}
\pdfoutput=1 
\title{Two-particle angular correlations of identified particles in pp and p--Pb collisions at LHC energies with ALICE}
\shorttitle{Angular correlation of identified particles in pp and p--Pb with ALICE at LHC}
\author{Daniela Ruggiano}
\institute{Warsaw University of Technology -- Warsaw, Poland}

\begin{document}
\maketitle
\vspace{-3.5cm}
\begin{abstract}
The two-particle angular correlations in the $\Delta$y,$\Delta\varphi$ space provide valuable insights into the properties of hadronization mechanisms and quark–gluon plasma properties. The correlation functions are influenced by several physical sources, including mini-jet correlations, Bose–Einstein quantum statistics, resonance decays, conservation of energy and momentum, and other factors.
Each correlation source has unique properties, and therefore each correlation function has a distinct form depending on transverse momentum and/or multiplicity. Previous results from angular correlation analysis of pp collisions at the LHC energies indicate an anticorrelation for pairs of baryons of the same sign in $\Delta\eta,\Delta\varphi$ space. This contradicts the predictions of Monte Carlo models, such as PYTHIA8 and EPOS. \\
This study aims to investigate this behavior by exploring the correlation functions of different charge combinations of the detected particles (specifically, $\rm{\pi^{\pm}}$, $\rm{K^{\pm}}$, and $\rm{p}$$\bar{\rm{p}}$) and multiplicity classes in the $\Delta$y$,\Delta\varphi$ space for pp and p--Pb collisions at LHC energies. In addition, the study includes a comparison of the results obtained from both collision systems.
\end{abstract}
\vspace{-1cm}
\section{Introduction}
The lack of a clear understanding of the production processes of one of the most abundant and fundamental particles in the universe, the proton, is one of the open and unresolved questions in the field of nuclear physics. Angular correlations are a powerful tool used to describe the mechanism of particle production in high-energy heavy-ion collisions. They are influenced by multiple effects such as (mini)jets, Bose--Einstein or Fermi--Dirac quantum statistics (in the case of identical bosons or fermions), resonance decays, Coulomb interactions, conservation of energy and momentum, and others, all of which contribute with some structure to the overall shape. Moreover, each correlation shows a unique behavior with its shape, and its individuality, as already observed in studies of proton--proton (pp) collisions at $\sqrt{s} = 7$ TeV by the ALICE Collaboration at LHC \cite{Adam_2017}. In particular, correlations of baryon--baryon pairs (combined with antibaryon--antibaryon pairs) in pp collisions exhibit a depletion, i.e., an anticorrelation, around ($\Delta\eta$,$\Delta\varphi) \sim (0,0)$. The origin of this phenomenon is still unclear. Several studies have been conducted on it, becoming a puzzle for many researchers \cite{Adam_2017, starcoll}. These studies have been extended to different baryons to investigate the shape of the correlation functions, which are related to the intrinsic nature of baryon production. None of the observed baryon--baryon correlations agree even qualitatively with the theoretical Monte Carlo models \cite{Adam_2017, DRuggiano}. In this work, which is an extension of the analysis of the pp collision at $\sqrt{s} = 13$ TeV \cite{DRuggiano}, another piece is proposed to be added to the puzzle. The study concerns the analysis of correlation functions in four different multiplicity classes in $\Delta$y,$\Delta\varphi$ in the proton--nucleus (p--Pb) collision at $\sqrt{s_{\rm NN}} = 5.02$ TeV. The results will be compared with the corresponding ones from pp collisions.

\section{Definition of two-particle correlation function}
In Ref.\cite{Adam_2017} the results were obtained using the so-called probability-ratio definition of the correlation function \cite{probability}, which experimentally is constructed as follows  
\begin{equation}
    C_{P}(\Delta y\Delta\varphi) = \frac{N_{pairs}^{mixed}}{N_{pairs}^{signal}} \frac{S(\Delta y\Delta\varphi)}{B(\Delta y\Delta\varphi)},
    \label{equation_prob}
\end{equation}
where S($\Delta$y$\Delta\varphi$) is the signal distribution and B($\Delta$y$\Delta\varphi$) is the background distribution. The signal distribution consists of pairs of particles from the same events, whereas the background distribution consists of pairs of particles from different events. It is normalized by the ratio of the number of pairs from the background and signal distributions. This definition has some limitations for studying correlations over different multiplicity classes due to the normalization factor that produces a trivial scaling of $1/N$, where N is the number of particles per collision \cite{DRuggiano}. This makes the comparison of different collision systems difficult because they are characterized by a large difference in multiplicity.
To avoid this issue, the rescaled two-particle cumulant definition is used for the first time in ALICE \cite{rescaled}. This new correlation function is defined as 
\begin{equation}
    C_{C}(\Delta y\Delta\varphi) = \frac{N_{av}}{\Delta y\Delta\varphi} (C_{P} - 1).
        \label{equation_cumulant}
\end{equation}
where $N_{av}$ is the average number of particles produced in the analyzed multiplicity classes.

\section{Analysis details}
The analysis is based on data recorded by the ALICE detector in p--Pb collisions at $\sqrt{s_{\rm NN}} = 5.02$ TeV in 2016.
To make the results of the two collision systems comparable,  the same event and track selection criteria were used in both analyses \cite{DRuggiano}. 
The angular correlation functions of the pairs of pions, kaons, and protons with the same and opposite sign were analyzed for four multiplicity classes corresponding to 0--20\% (highest multiplicities), 20--40\%, 40--70\%, and 70--100\% (lowest multiplicities) of the total interaction cross-section. The data sample consists of $\sim 9.05 \cdot 10^{8}$  minimum bias events \cite{MBiasTrigger}.
The angular acceptance for this analysis is $|y| < 0.5$ covering the full azimuthal angle. The analysis is based on reconstructed tracks in the following transverse momentum ranges: $0.2 < p_{\rm{T}} < 2.5$ GeV/$c$ for pions and $0.5 < p_{\rm{T}} < 2.5$ GeV/$c$ for kaons and protons.
Particle identification (PID) of pions, kaons, and protons is performed track by track using information from the TPC and TOF detectors \cite{detectorsdescription}. 
It is based on $N_{\sigma}$ method, where $N_{\sigma}$ is the number of standard deviations of the Gaussian distribution around the theoretical signal. For $p_{\rm T} < 0.5$ GeV/$c$, only the TPC information is used. 

\section{Results}
\subsection{Correlation functions in p--Pb}
\begin{figure}[h!]
    \centering
    \includegraphics[width=4.4cm]{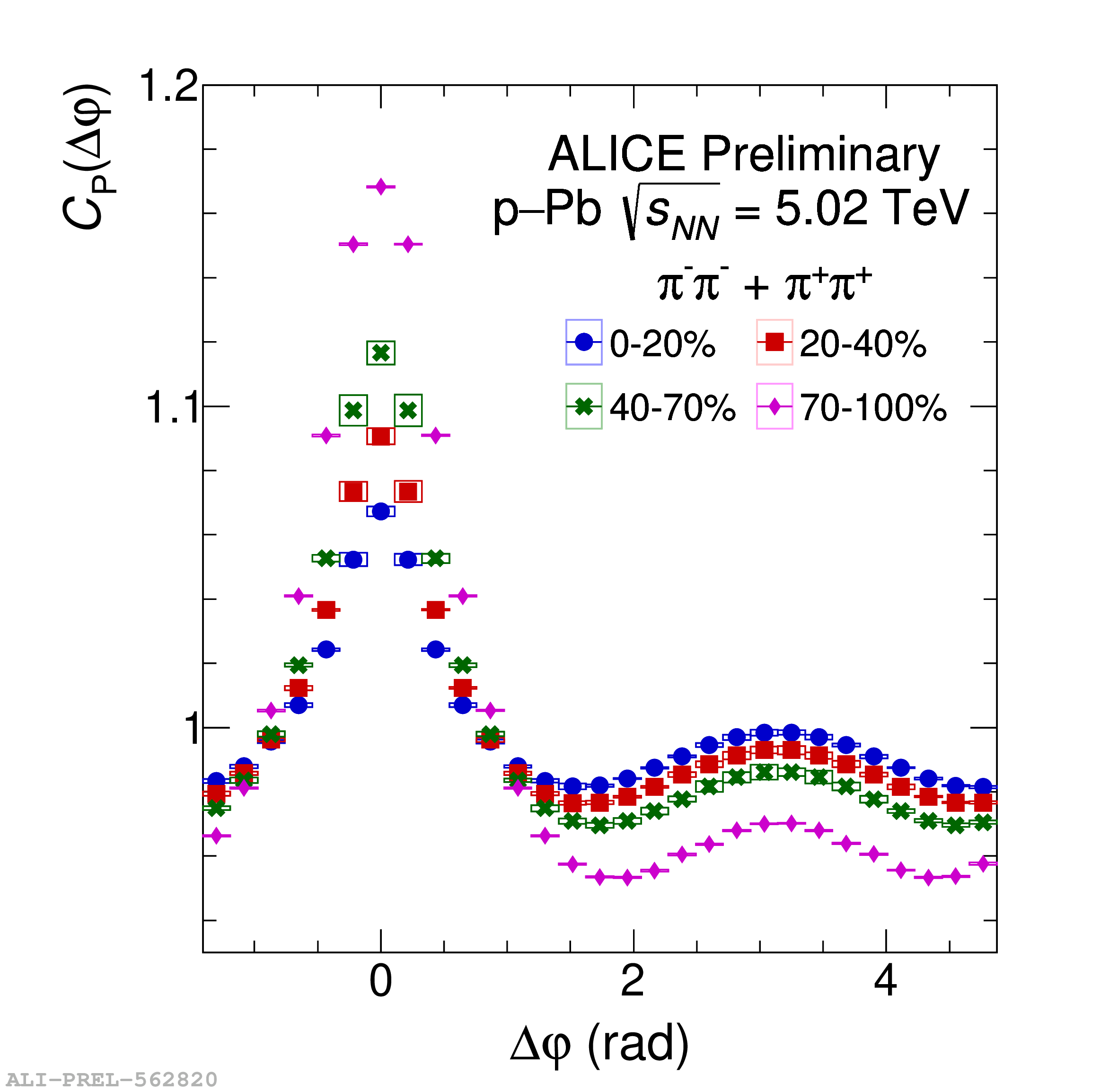}
    \includegraphics[width=4.4cm]{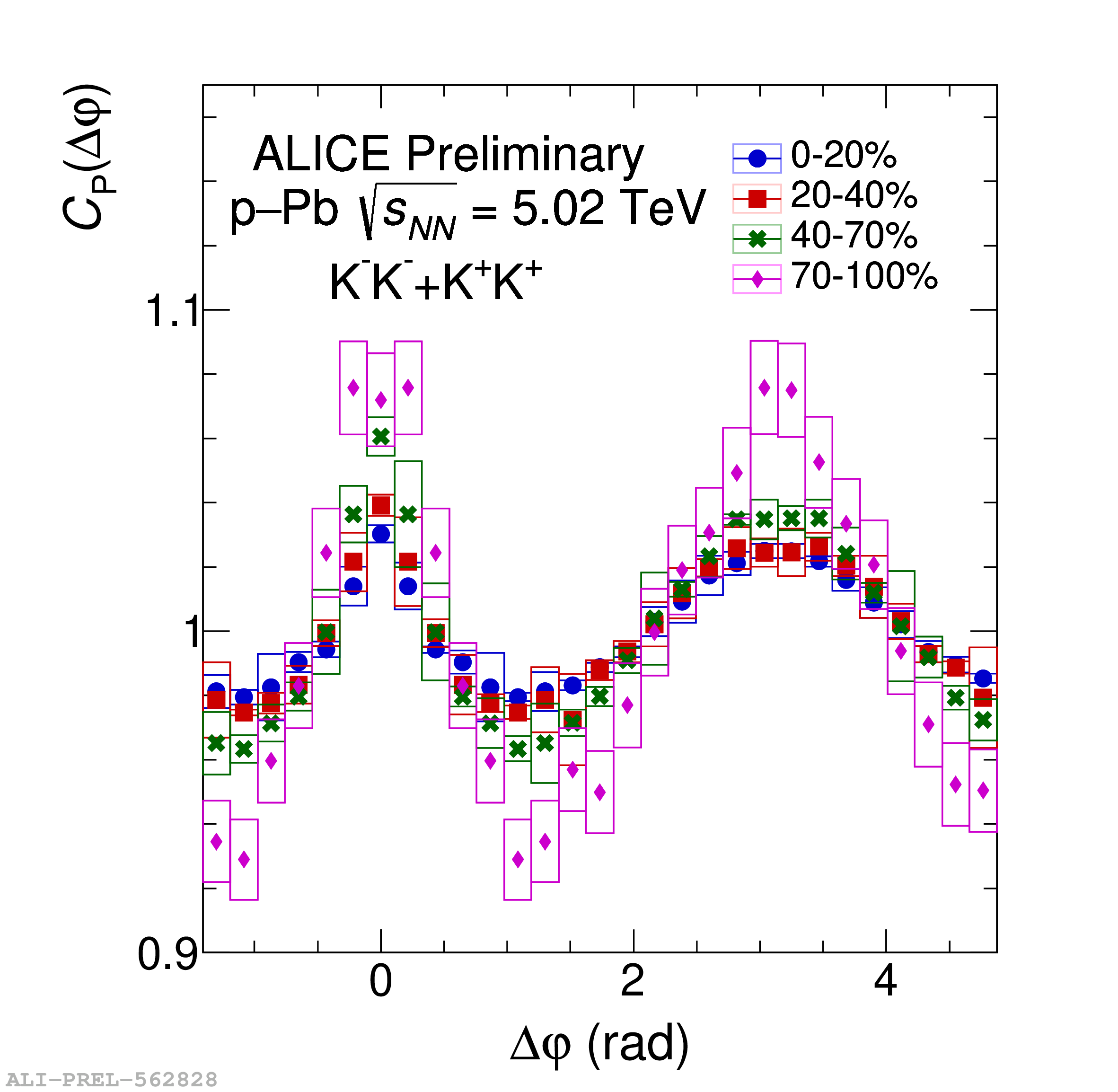}
    \includegraphics[width=4.4cm]{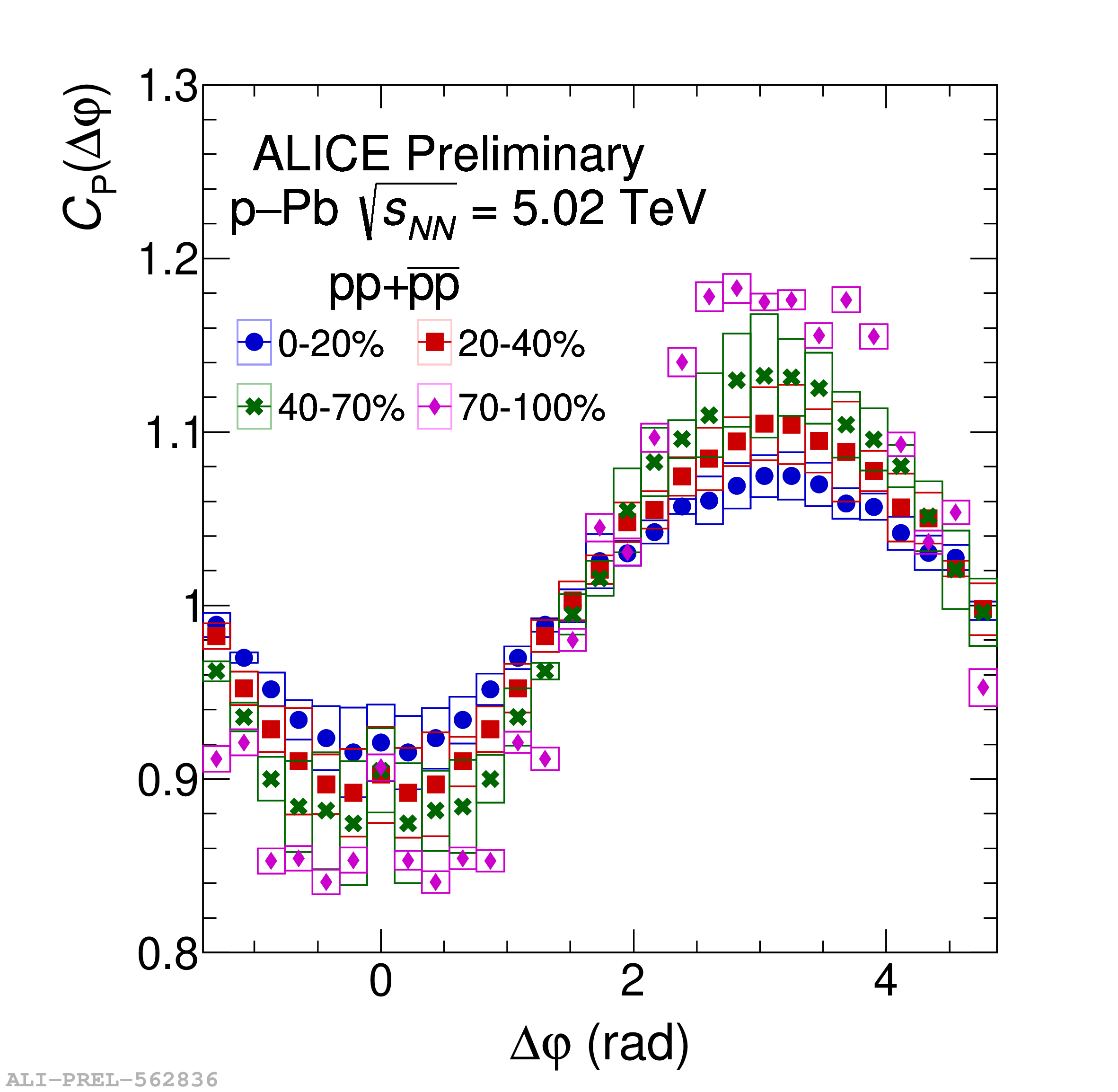}
    \includegraphics[width=4.4cm]{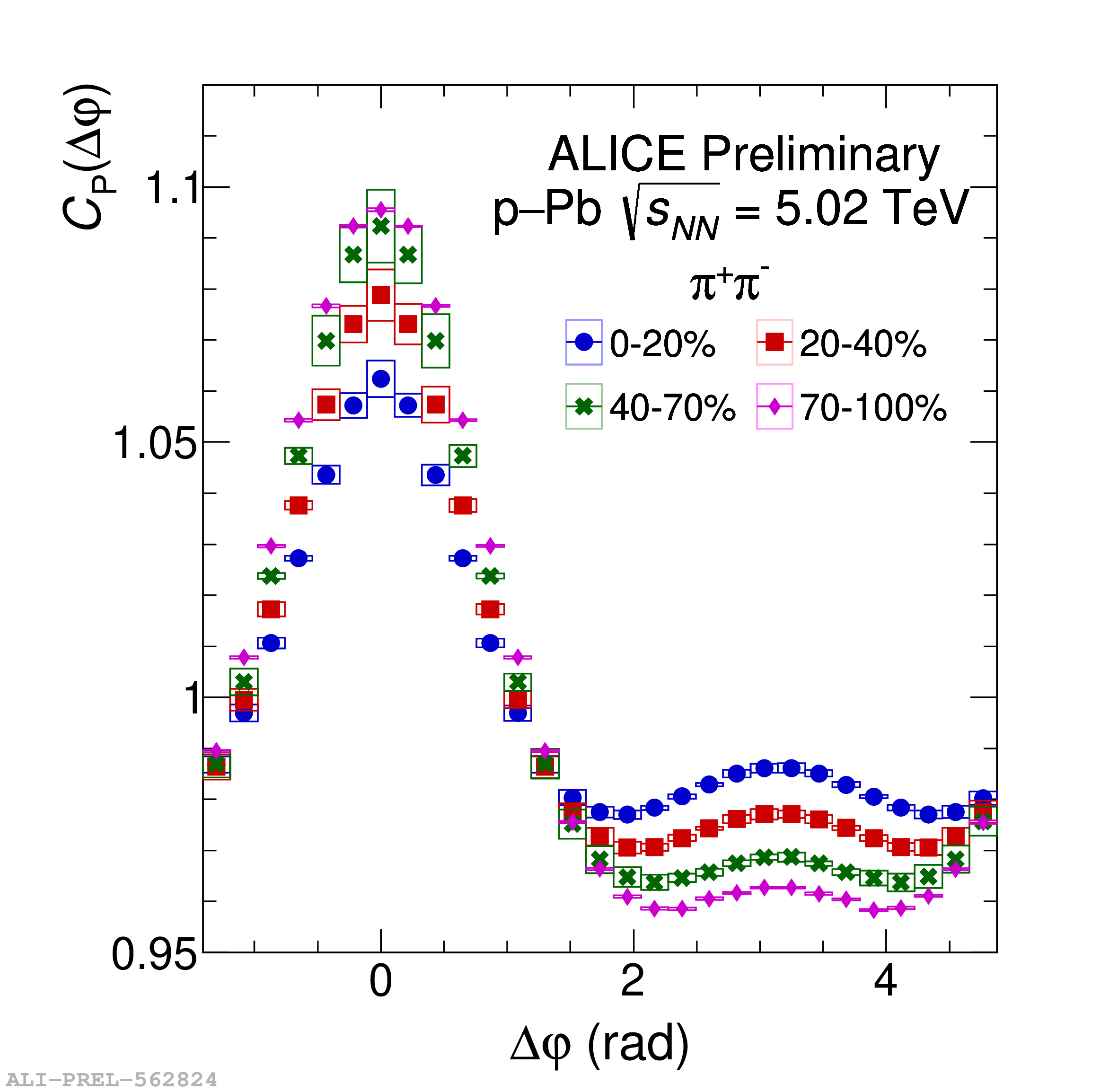}
    \includegraphics[width=4.4cm]{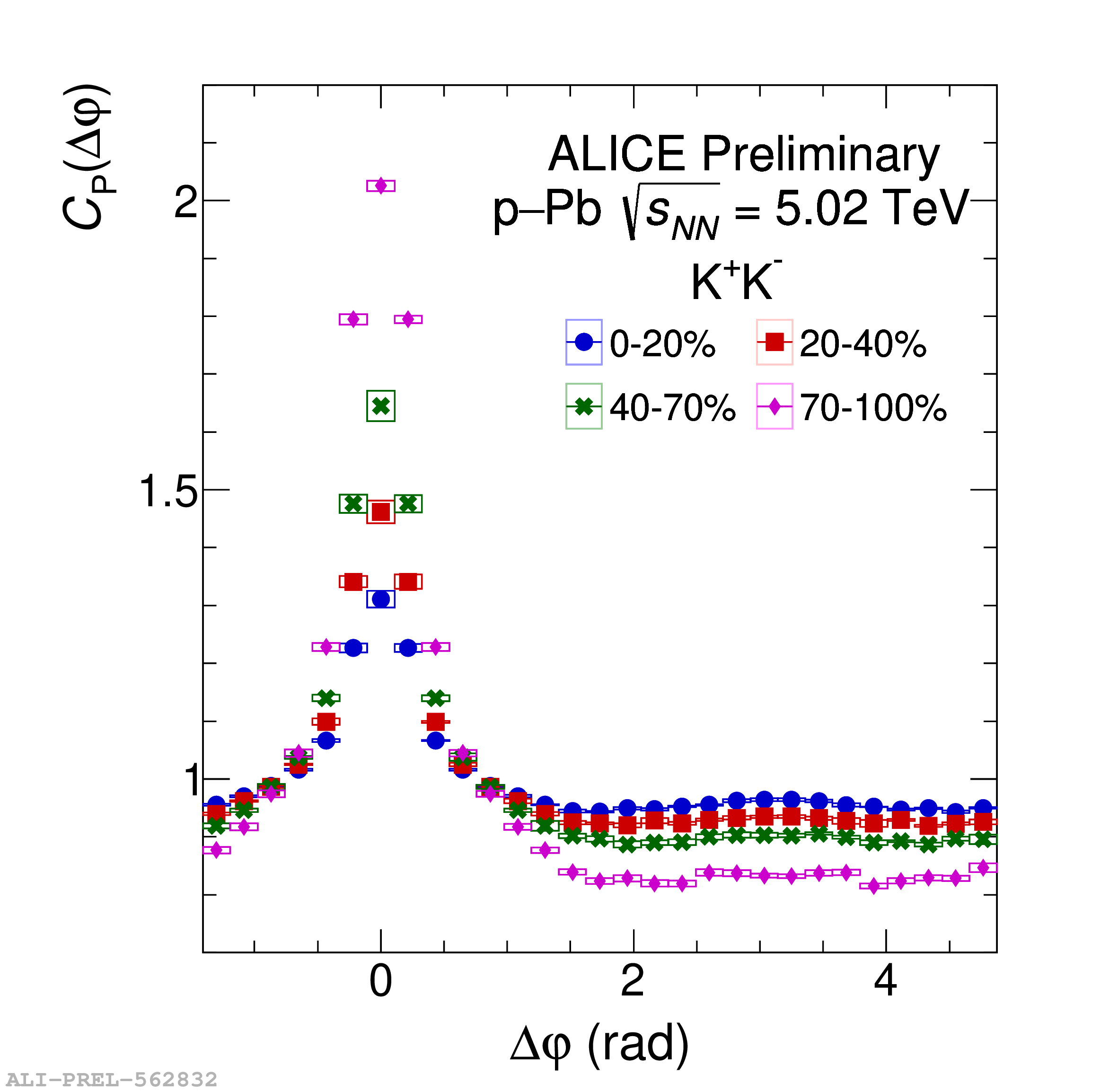}
    \includegraphics[width=4.4cm]{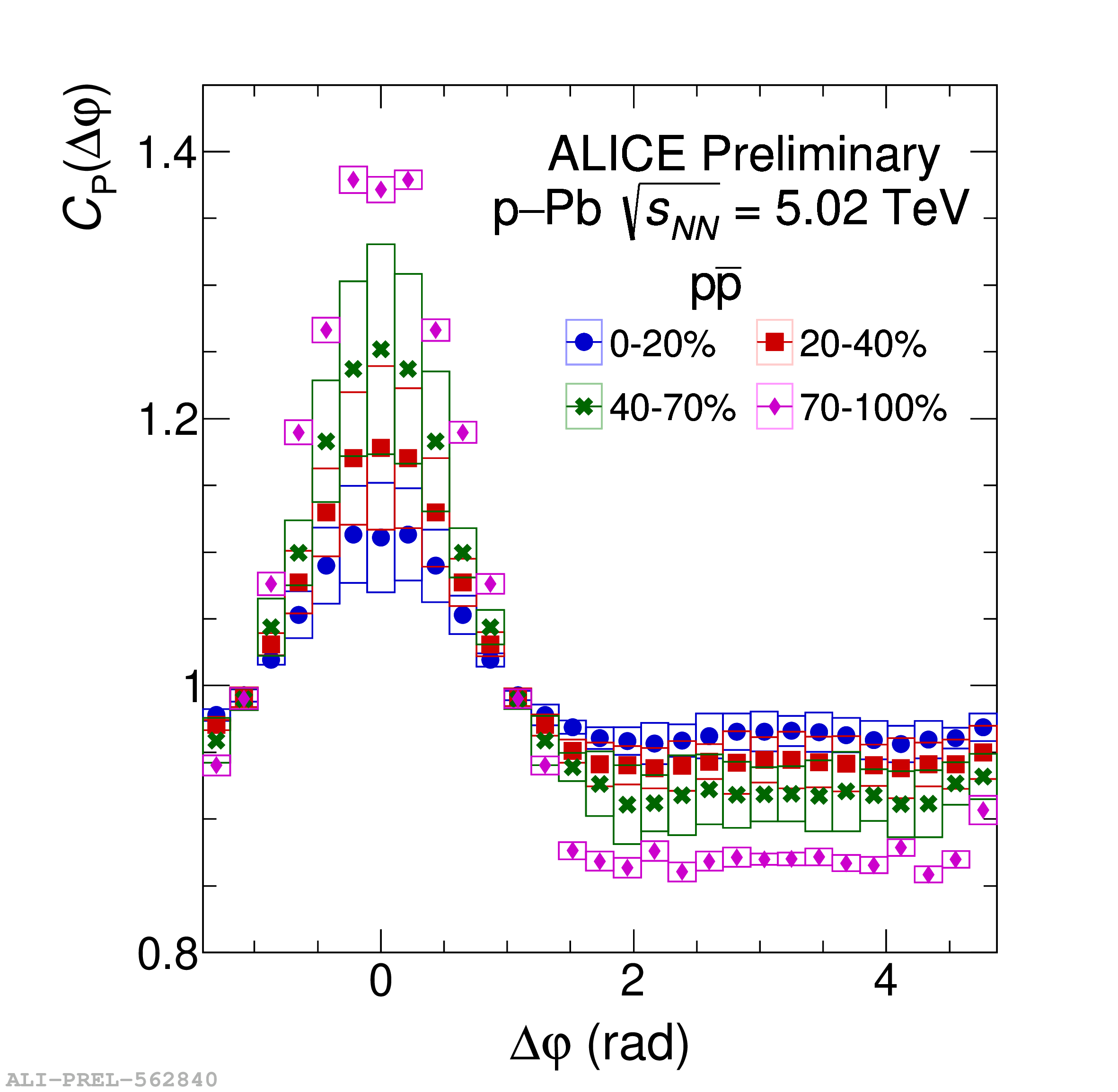}
    \caption{The $\Delta\varphi$ projection of correlation functions using the probability ratio definition in p--Pb collisions at $\sqrt{s_{\rm NN}} = 5.02$ TeV for four multiplicity classes. Particles with like signs are depicted on the top panels, while those with unlike signs are shown on the bottom panels.}
    \label{projection_probratio}
\end{figure}

\begin{figure}[h!]
    \centering
    \includegraphics[width=4.4cm]{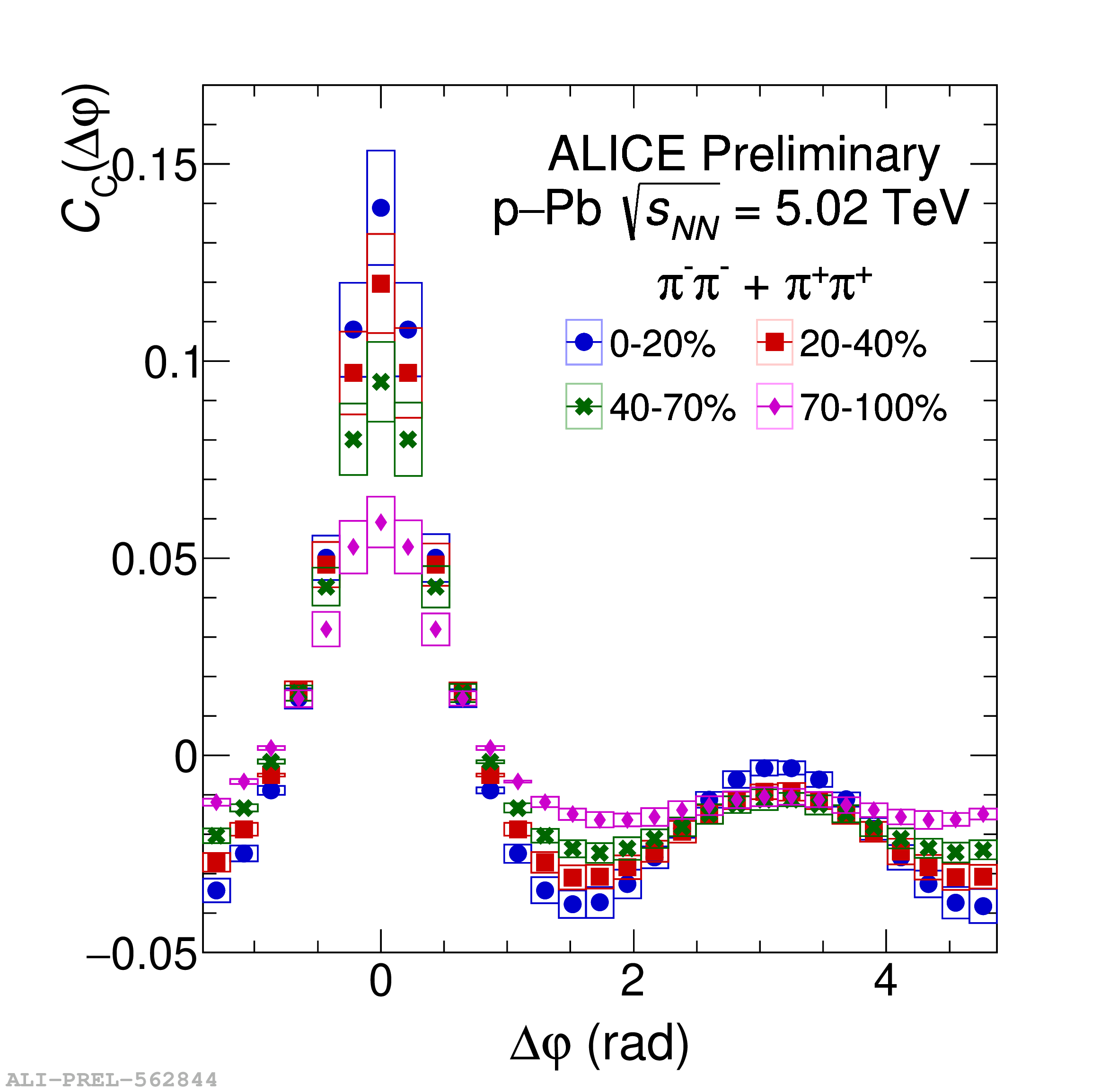}
    \includegraphics[width=4.4cm]{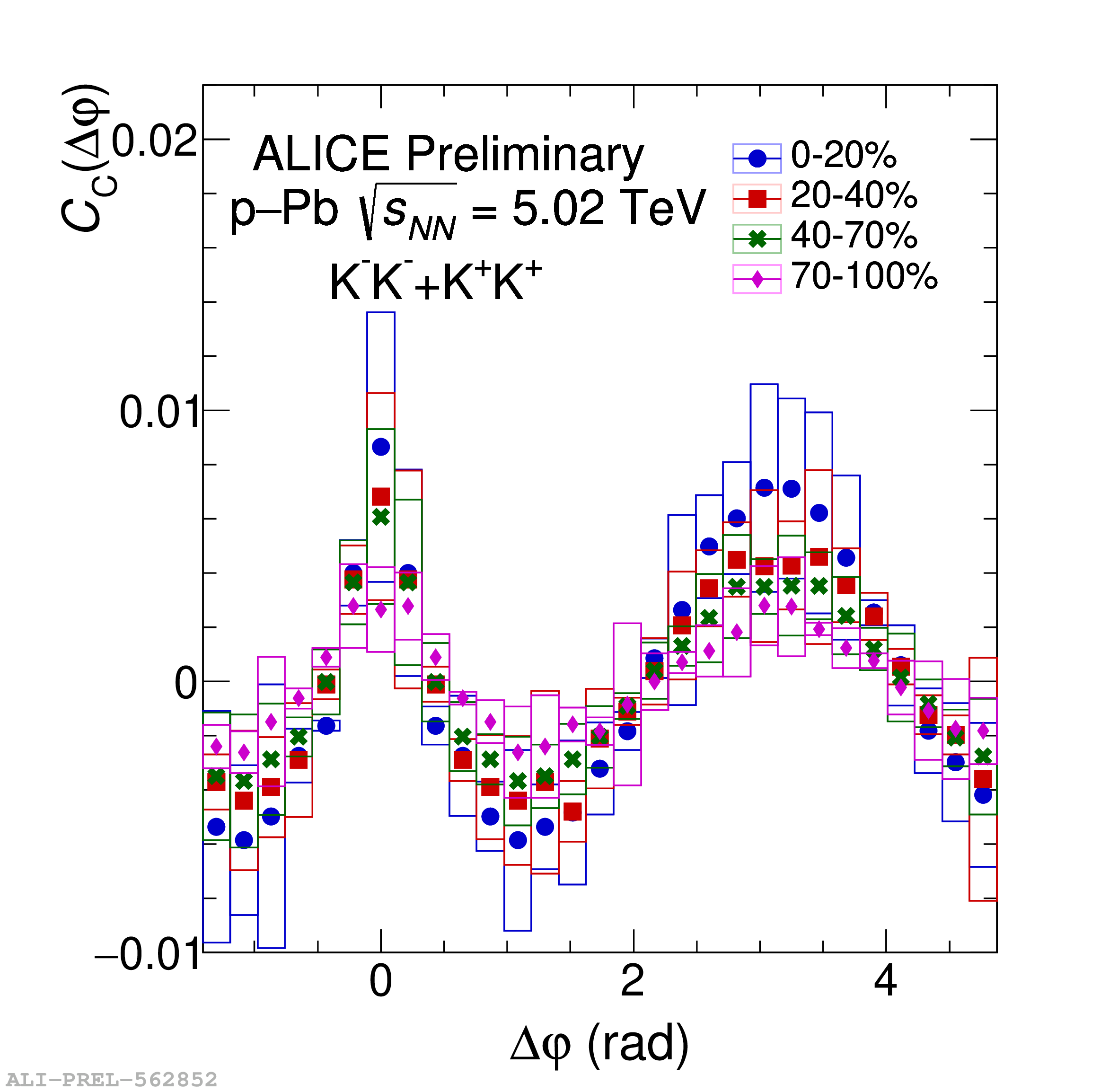}
    \includegraphics[width=4.4cm]{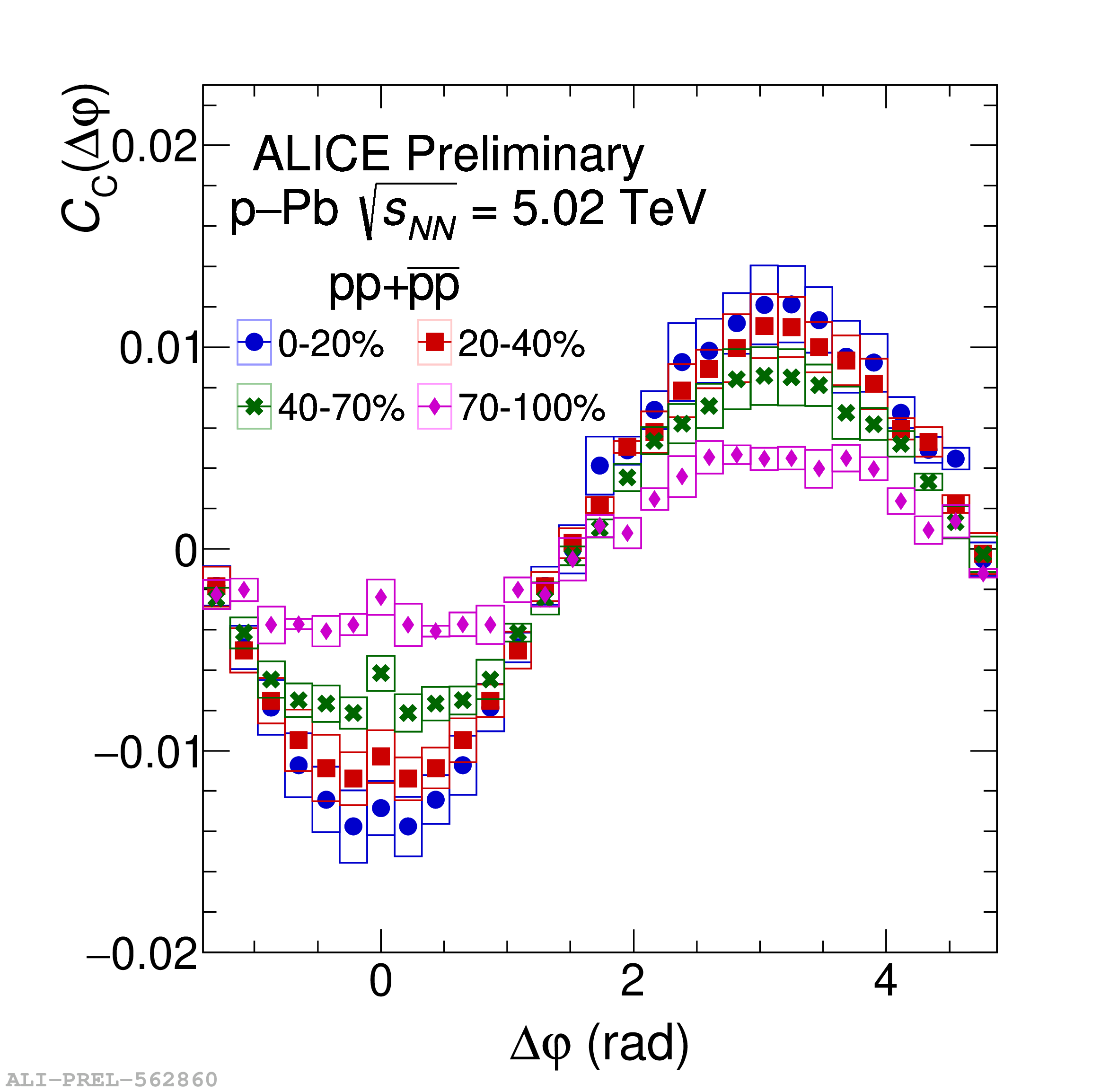}
    \includegraphics[width=4.4cm]{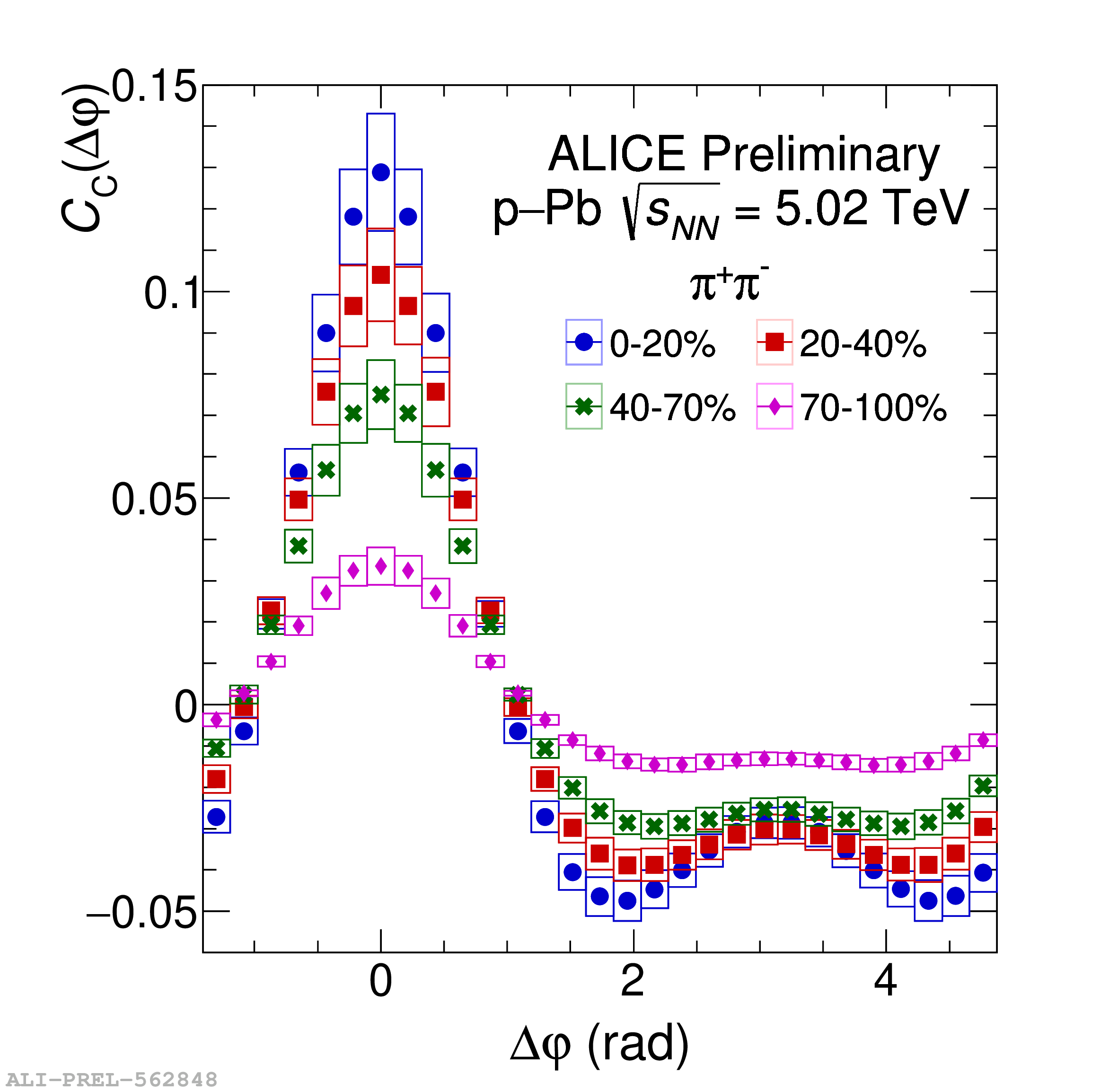}
    \includegraphics[width=4.4cm]{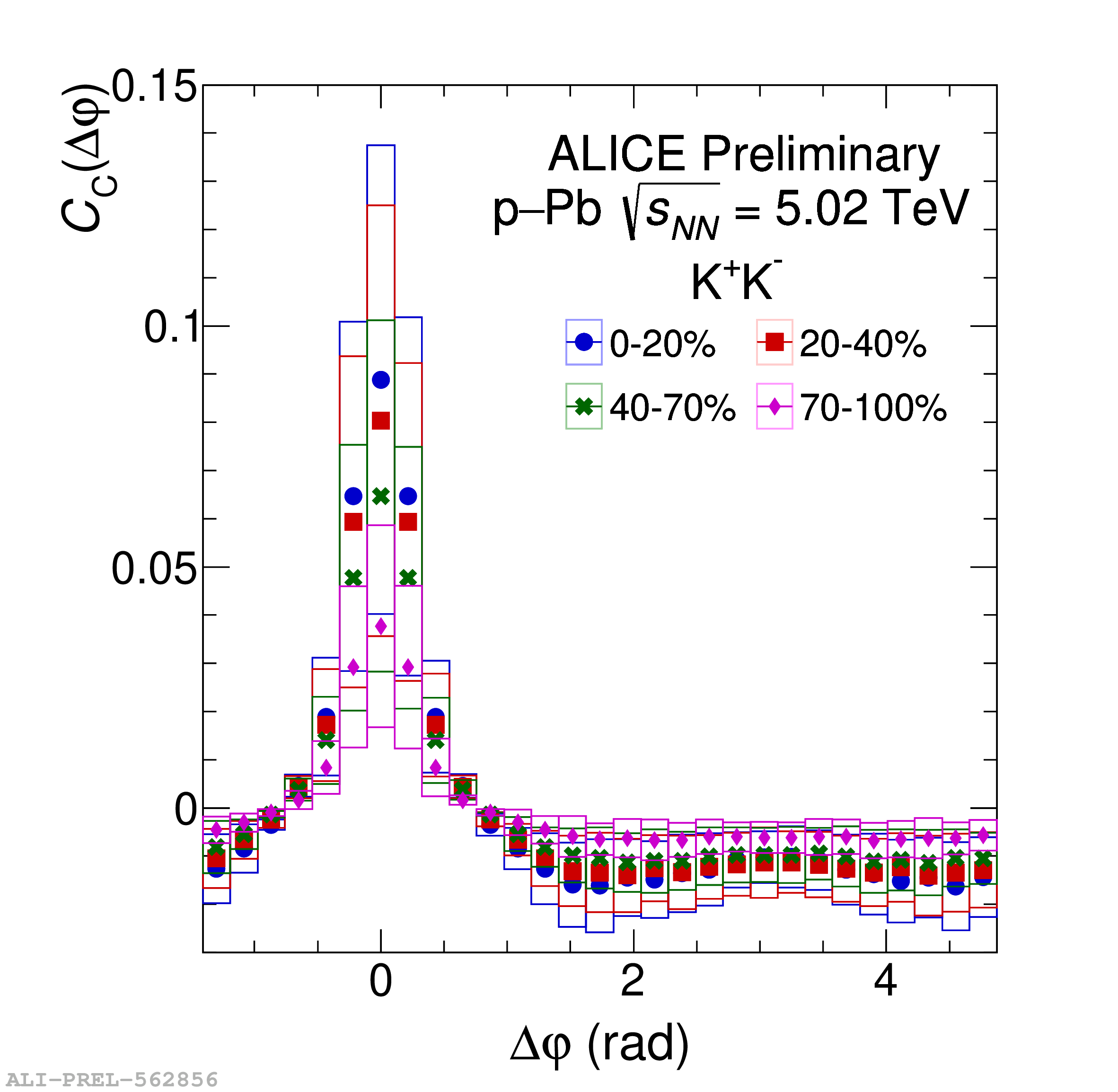}
    \includegraphics[width=4.4cm]{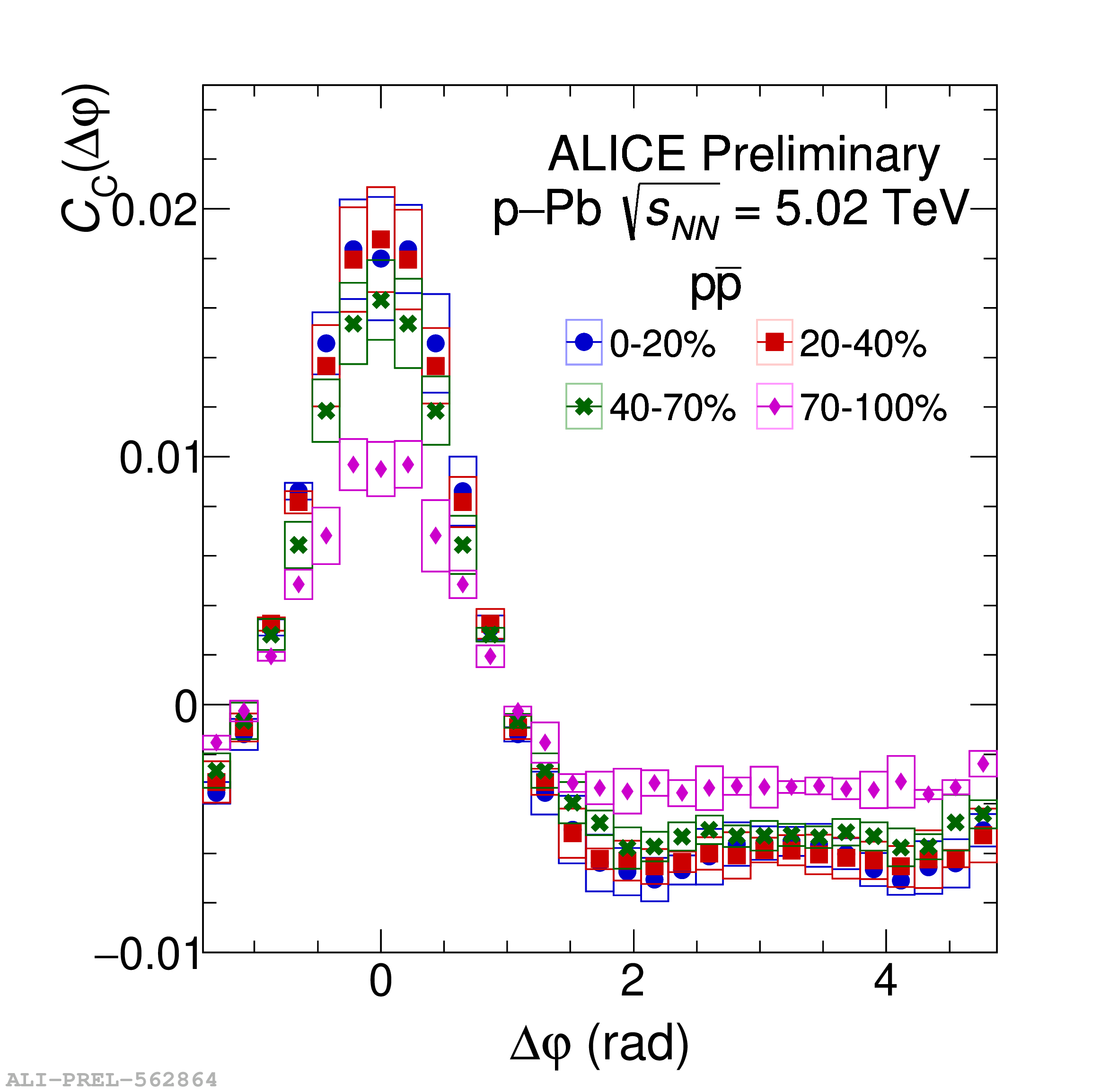}
    \caption{The $\Delta\varphi$ projection of correlation functions using the rescaled two-particle cumulant definition in p--Pb collisions at $\sqrt{s_{\rm NN}} = 5.02$ TeV for four multiplicity classes. Particles with like signs are depicted in the top panels, while those with unlike signs are shown on the bottom panels.}
    \label{projection_rescaled}
\end{figure}

Figures \ref{projection_probratio} and \ref{projection_rescaled} show the correlation functions for pion, kaon, and proton pairs measured in the p--Pb collisions at $\sqrt{s_{\rm NN}}$ = 5.02 TeV using the probability-ratio definition and the rescaled two-particle cumulants in the four multiplicity classes, respectively. Using the definition of probability ratio, the correlation functions are a convolution of physical phenomena plus the trivial scaling over different multiplicity classes. 
Indeed, there is an increase toward lower multiplicity classes due to the 1/N scaling. In contrast, the rescaled two-particle cumulant shows an opposite trend, that is, an increase in the correlation function with higher multiplicity. Therefore, by removing the 1/N factor, it is possible to make a more detailed study of the various physical phenomena that define the correlation functions.

\subsection{Comparison of $\Delta y,\Delta\varphi$ correlation functions between pp and p--Pb}
\begin{figure}[h!]
    \centering
    \includegraphics[width=4.4cm]{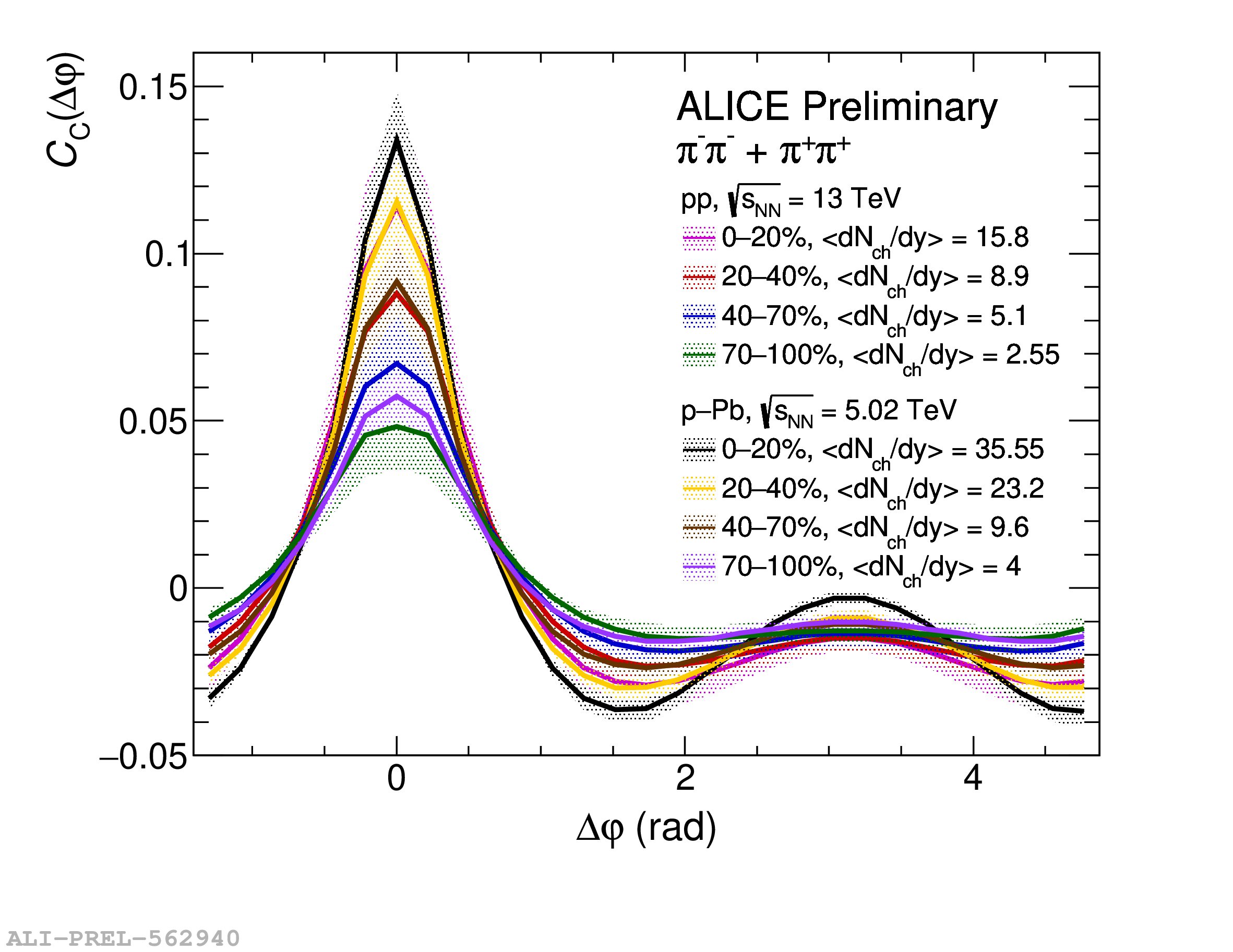}
    \includegraphics[width=4.4cm]{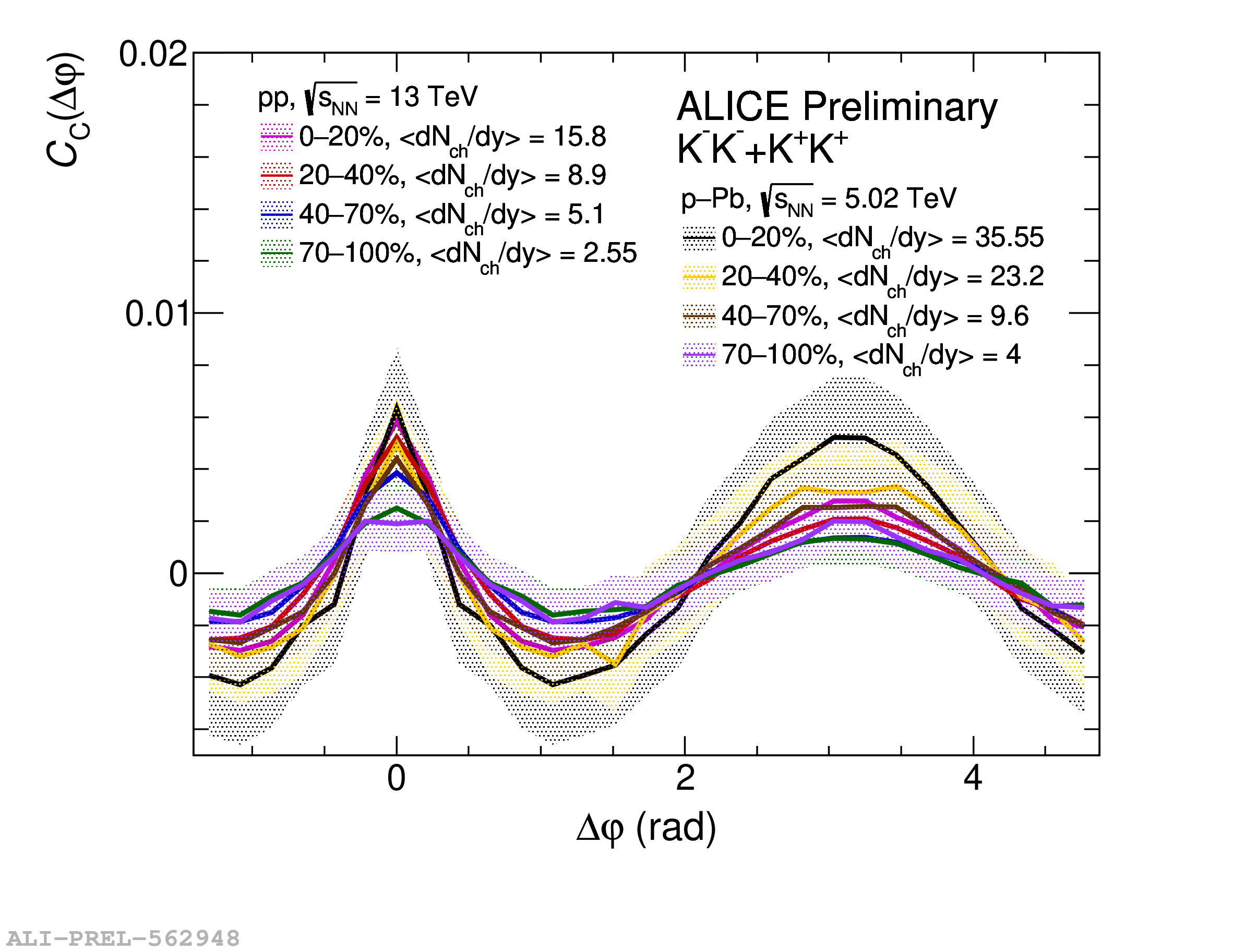}
    \includegraphics[width=4.4cm]{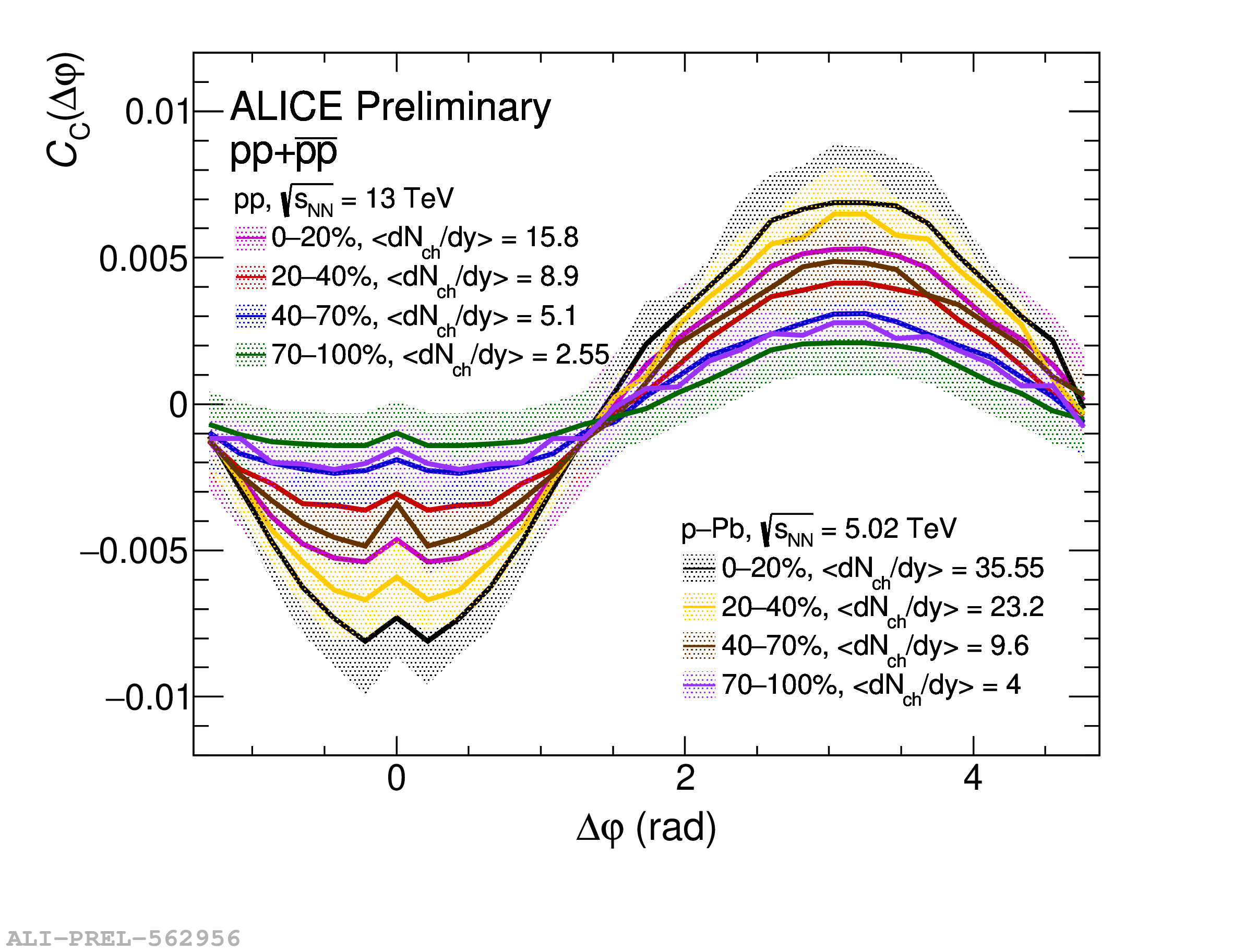}
    \includegraphics[width=4.4cm]{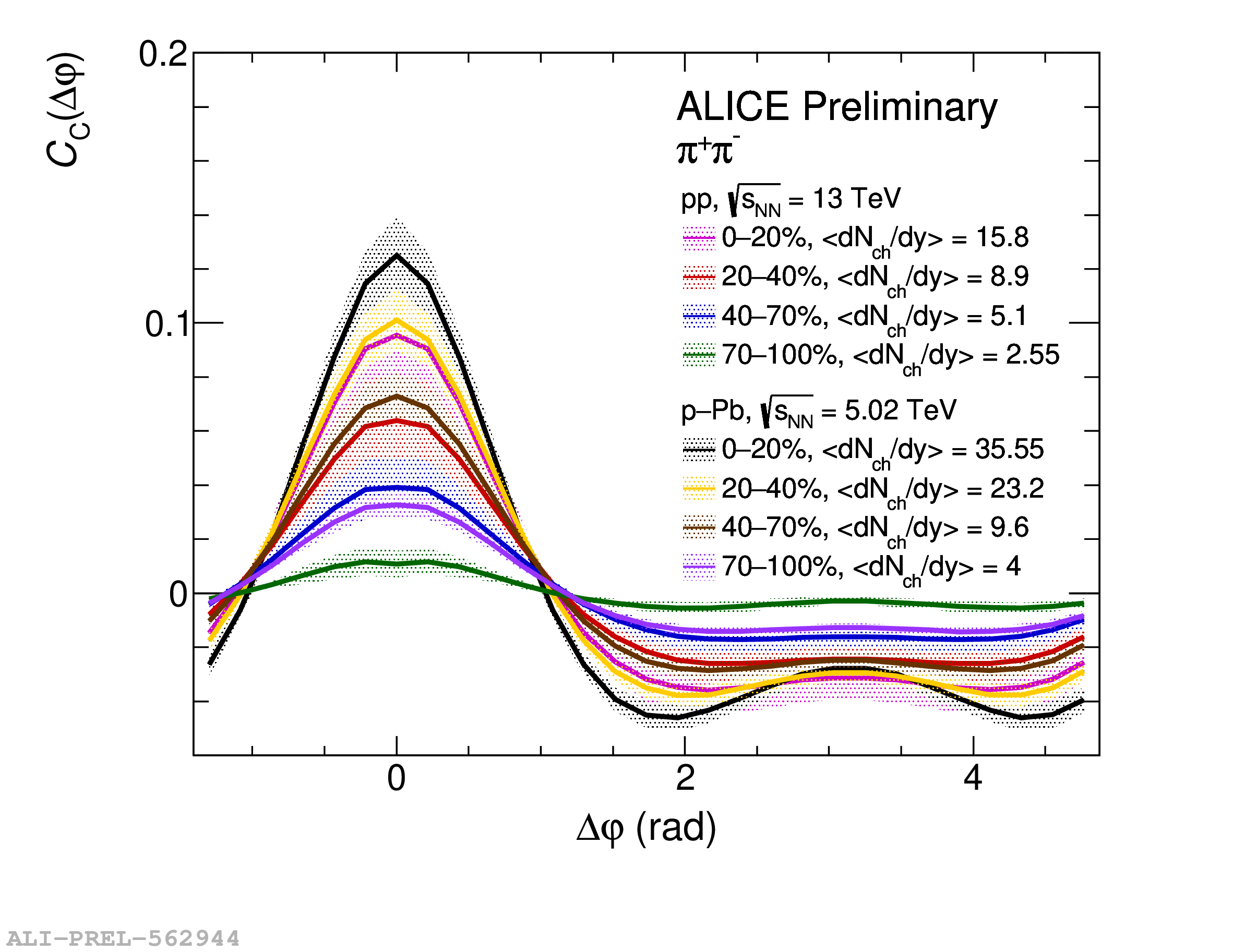}
    \includegraphics[width=4.4cm]{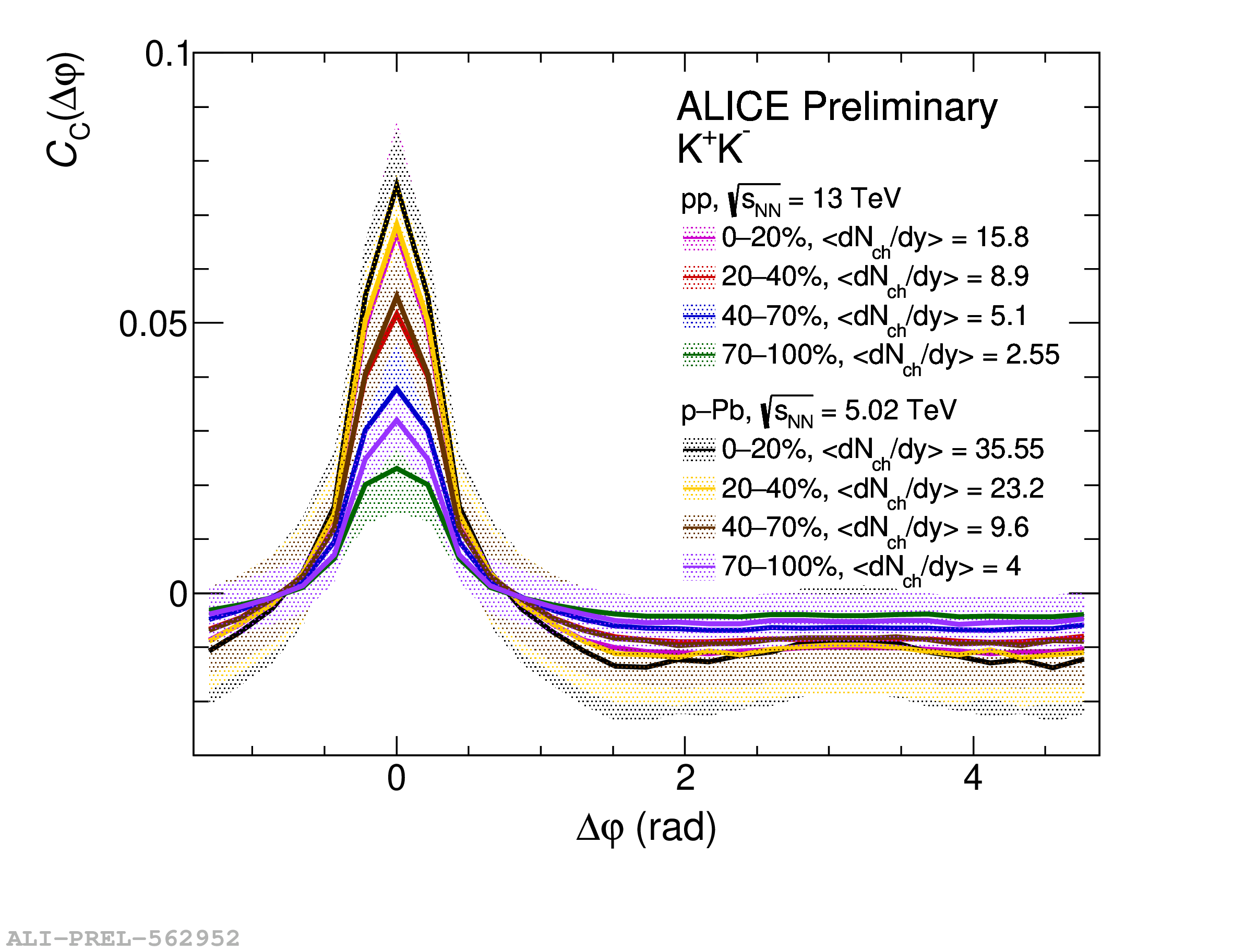}
    \includegraphics[width=4.4cm]{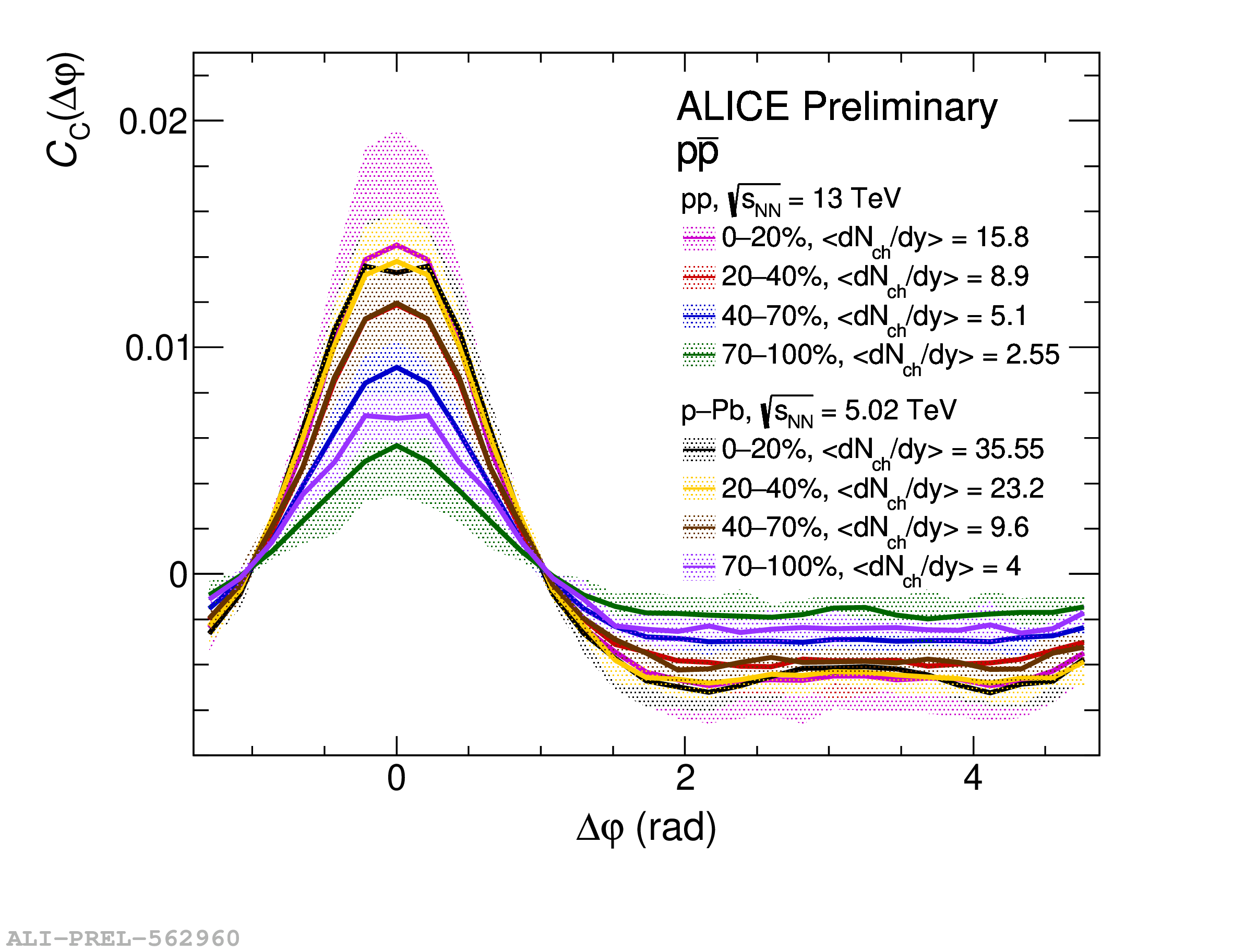}
    \caption{Multiplicity dependence of the correlation functions measured using the rescaled two-particle cumulant from pp collisions at $\sqrt{s} = 13$ TeV and p--Pb collisions at  $\sqrt{s_{\rm NN}} = 5.02$ TeV.}
    \label{comparison_pp_pPb}
\end{figure}

The results obtained in pp collisions \cite{DRuggiano} are compared with those for p--Pb collisions using the definition of rescaled two-particle cumulant in Fig. \ref{comparison_pp_pPb}.
In p--Pb collisions, the correlation functions are stronger than in pp collisions due to the higher multiplicities, which is also evident in the baryon–baryon correlations, where an anticorrelation is observed. 
In the case of proton--antiproton correlation, it is observed that for the highest multiplicity class (0--20\%) the correlation function for pp collisions appears to be stronger than that observed in p--Pb collisions. This is due to the annihilation phenomenon visible in the correlation function represented by the dashed black line, which shows an anti-peak at $\Delta\varphi$ = 0. This physical effect has been studied previously using the femtoscopy method with Monte Carlo simulations of PYTHIA 8 from pp collisions at $\sqrt{s} = 13$ TeV \cite{Unfolding}. The observed depletion due to the strong interaction is qualitatively in agreement with experimental observations of the STAR Collaboration in Au--Au collisions and model comparisons, which confirm that the anticorrelation in $\rm{p}$$\bar{\rm{p}}$ is caused by the annihilation phenomenon \cite{AnnichilationSTAR}. It is therefore difficult to make a comparison with the correlation function in pp collisions in the 0--20\% multiplicity class.
\subsection{Multiplicity comparison between pp and p--Pb}
The ${\rm d}N_{ch}/\rm{d}\eta$ values for all multiplicity classes in pp and p--Pb collisions included in this analysis are taken from Refs. \cite{dNchdetavaluespp, dNchdetavaluespPb} and are listed in Tab. \ref{table_dNch_deta}.
\begin{table}[ht]
\centering
\small{
\begin{tabular}{|c|c|c|}
\hline
${\rm d}N_{ch}/{\rm d}\eta$     & pp   & p--Pb \\ \hline
 0-20\%   & 15.8 & 35.55 \\ \hline
 20-40\%  & 8.9  & 23.2  \\ \hline
 40-70\%  & 5.1  & 9.6   \\ \hline
 70-100\% & 2.55 & 4     \\ \hline
\end{tabular}
\caption{The ${\rm d}N_{ch}/{\rm d}\eta$ values for the investigated multiplicity classes in pp and p--Pb collisions.}
\label{table_dNch_deta}
}
\end{table}

\begin{figure}[h!]
    \centering
    \includegraphics[width=4.4cm]{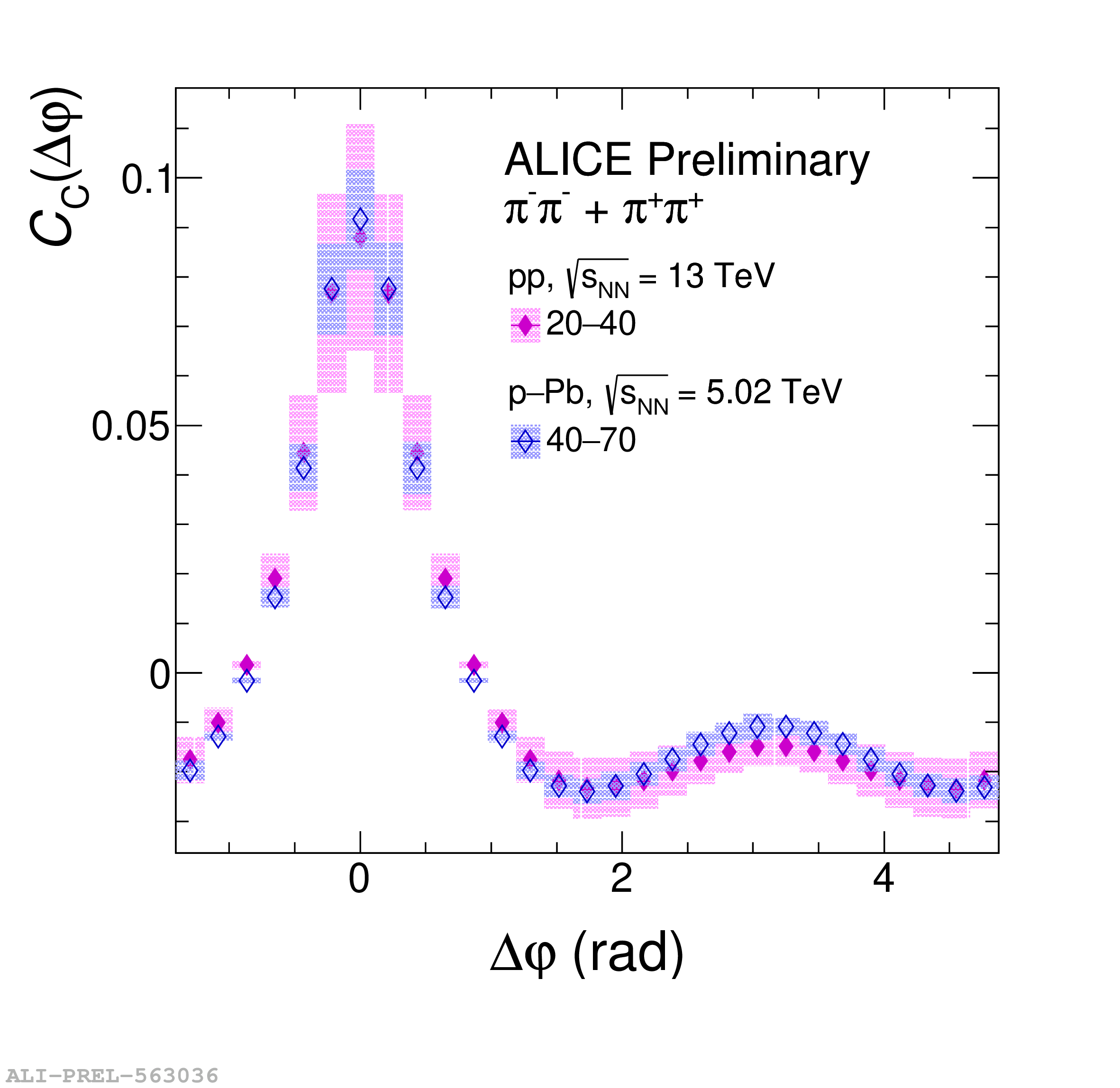}
    \includegraphics[width=4.4cm]{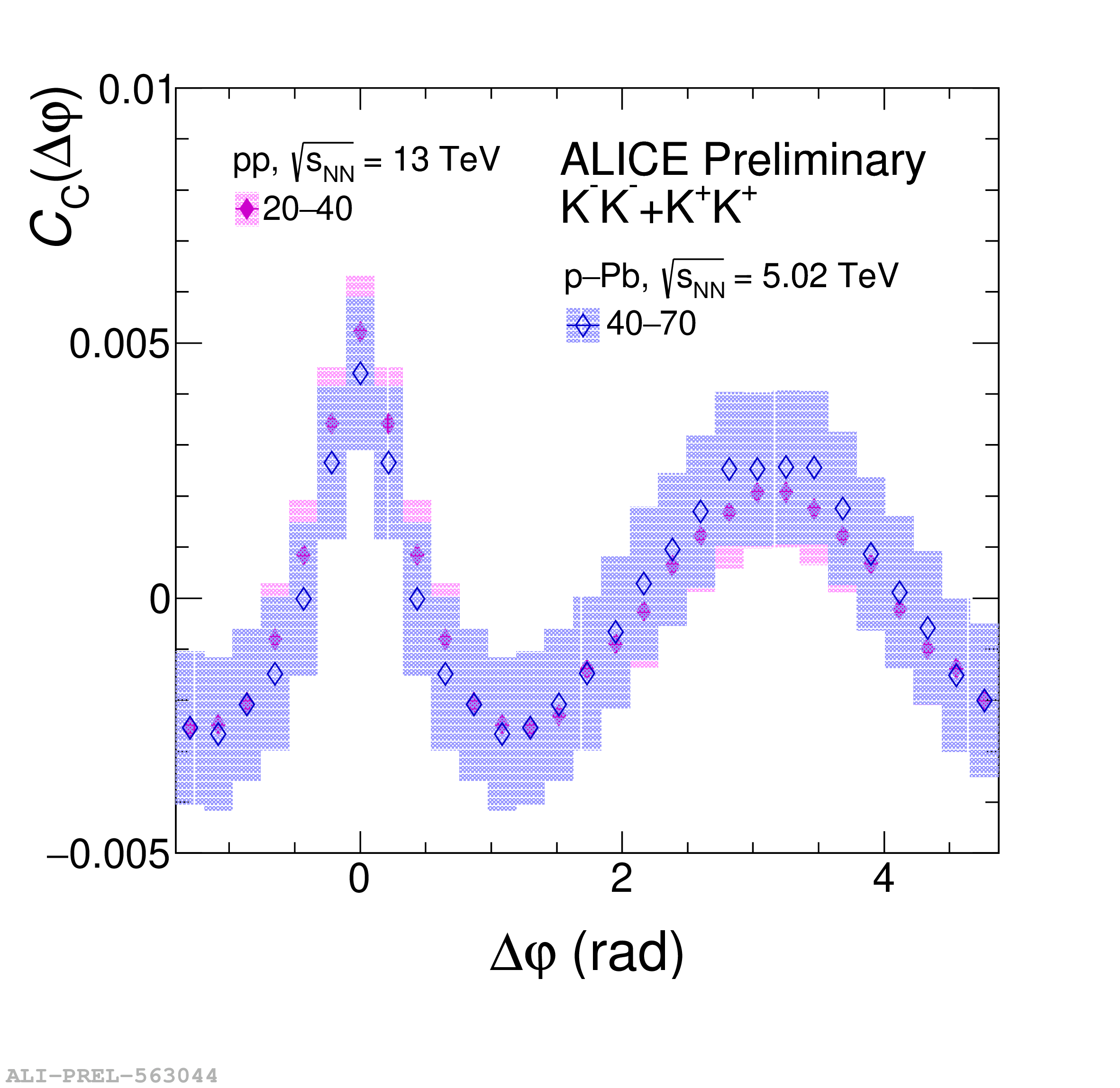}
    \includegraphics[width=4.4cm]{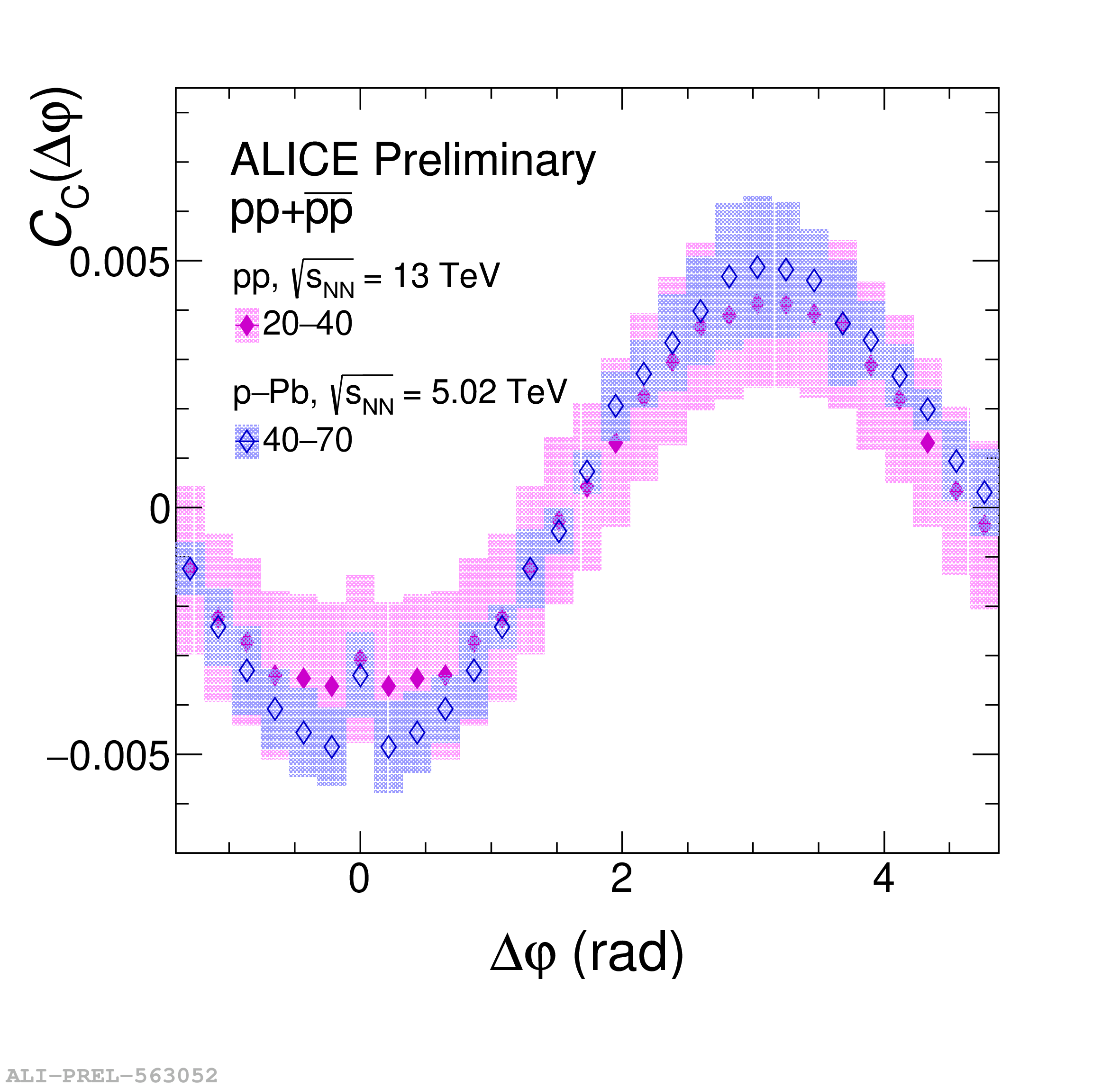}
    \includegraphics[width=4.4cm]{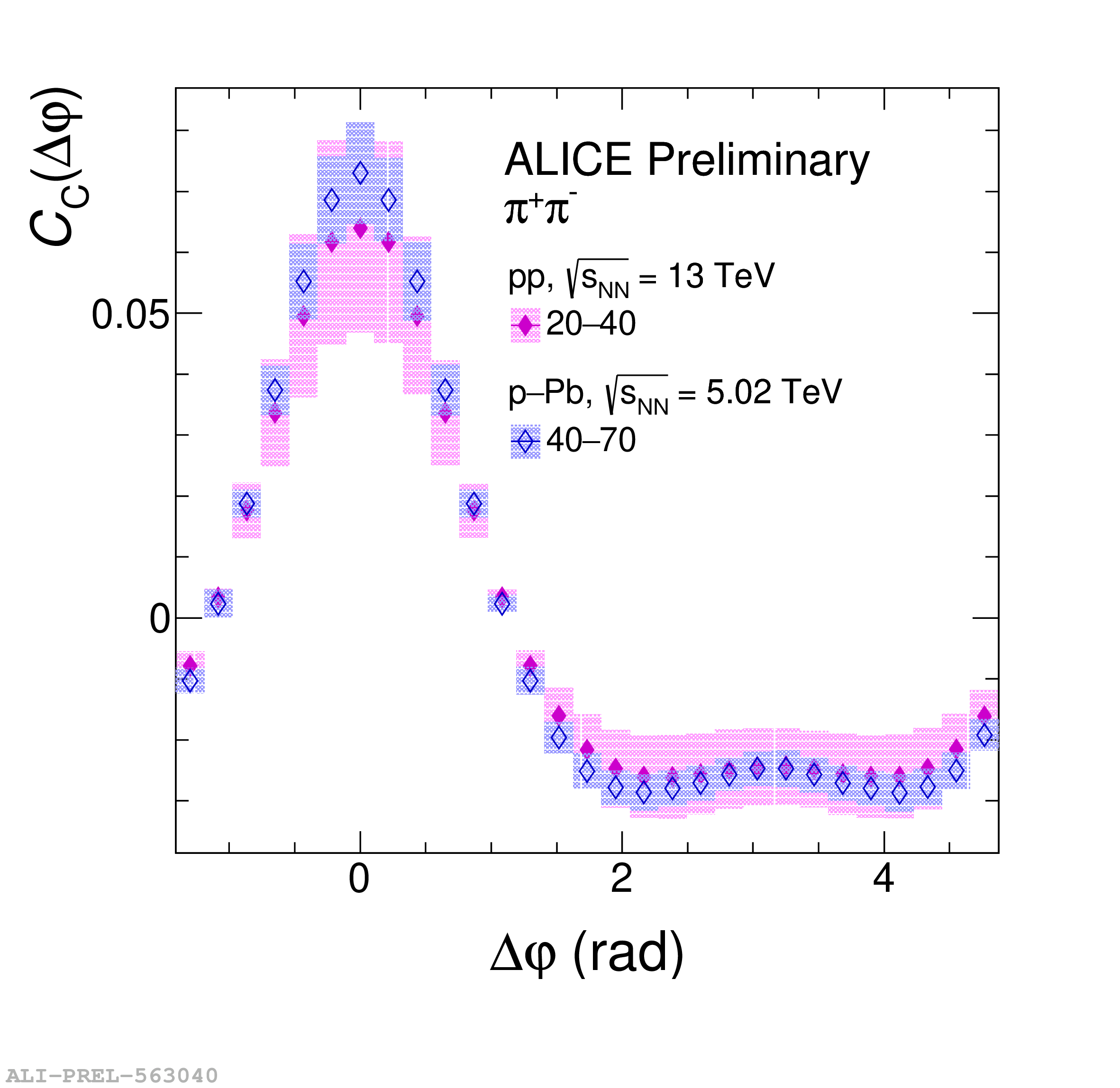}
    \includegraphics[width=4.4cm]{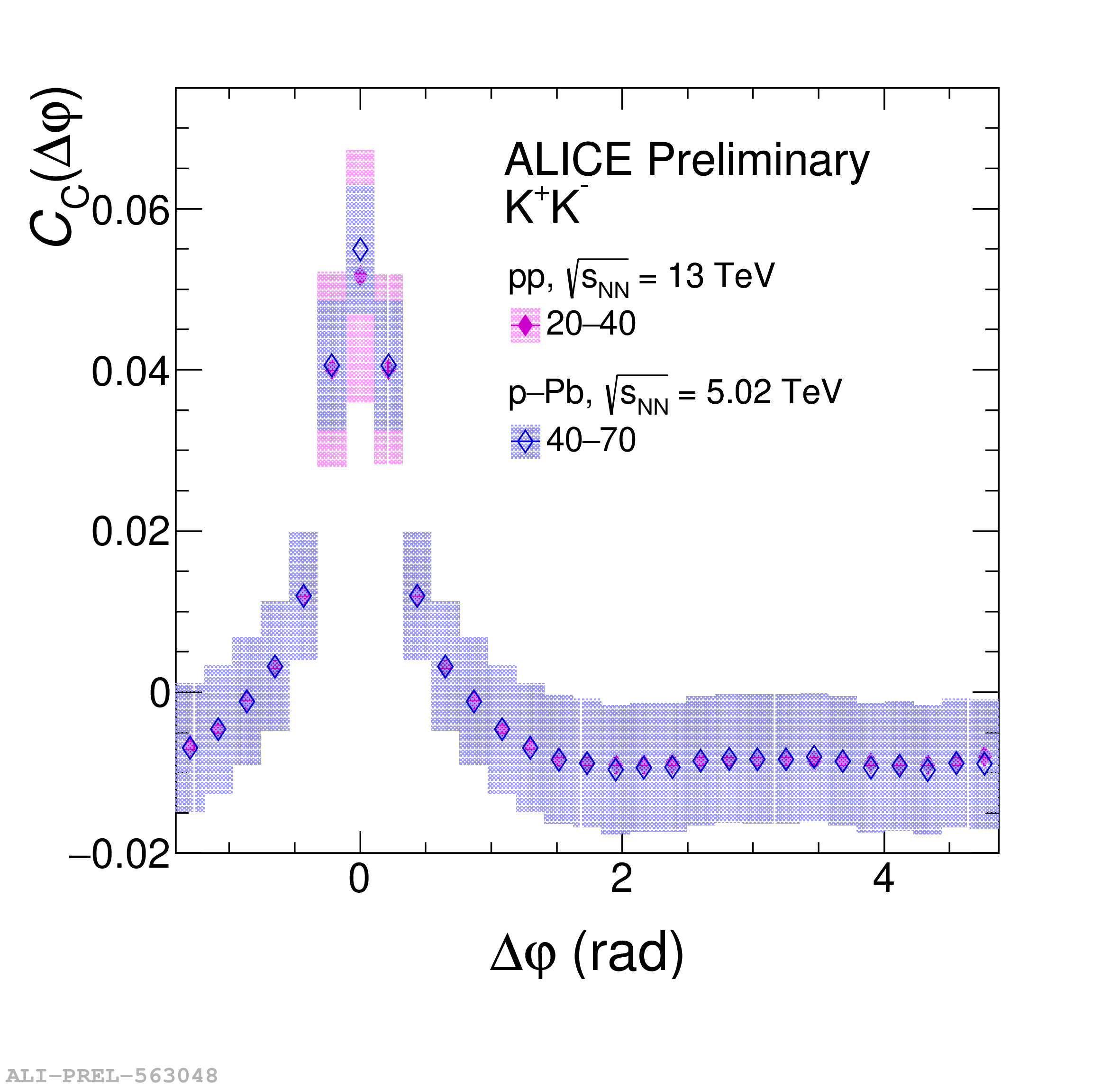}
    \includegraphics[width=4.4cm]{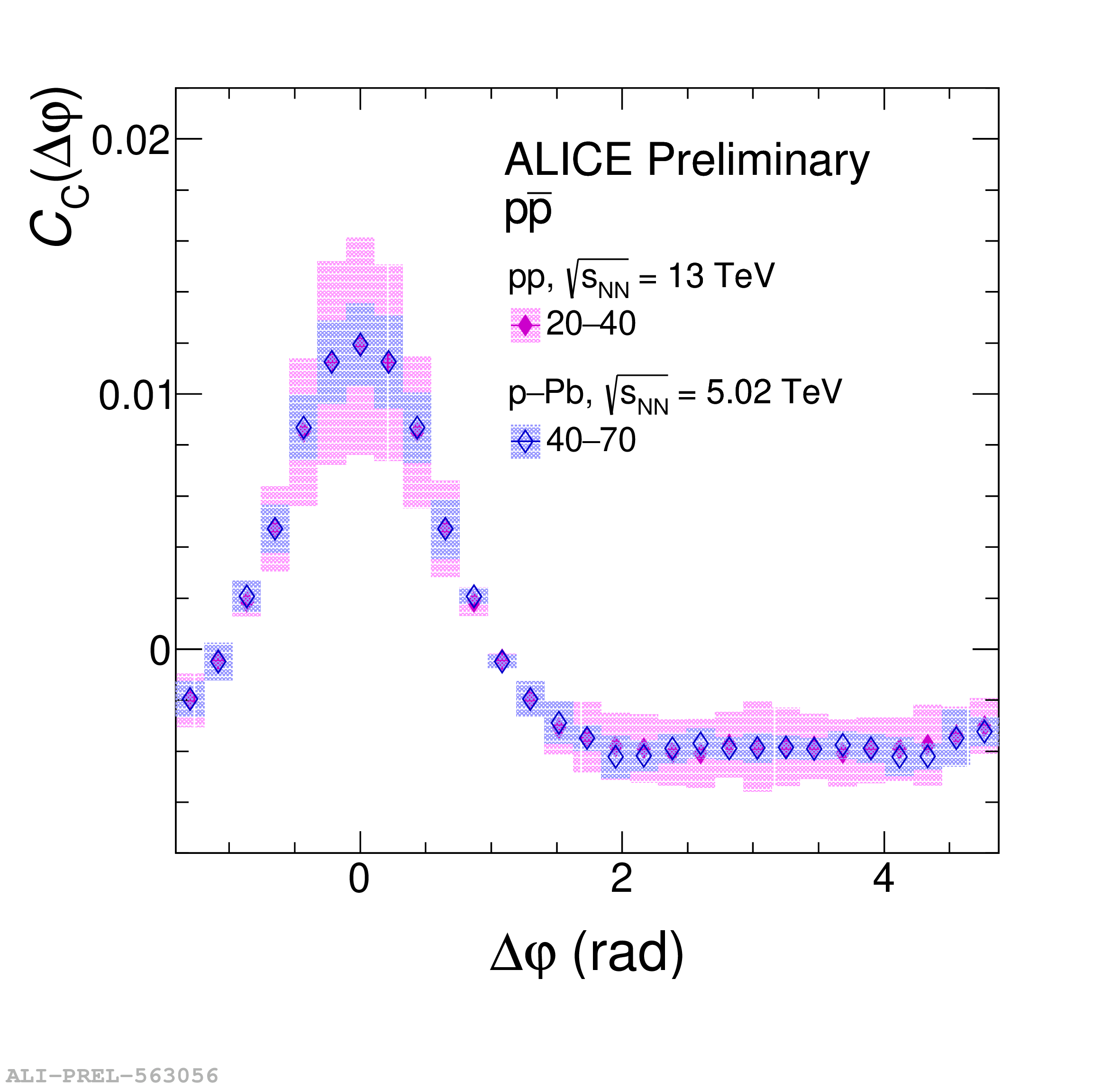}
    \caption{The comparison of the correlation functions between pp at $\sqrt{s} = 13$ TeV and p--Pb at $\sqrt{s_{\rm NN}} = 5.02$ TeV for the similar multiplicity classes using the rescaled two-particle cumulant definition.}
    \label{ComparisonMultiplicities}
\end{figure}
Based on these values, a rigorous comparison of the correlation functions is made between the closest multiplicities, i.e., 20--40\% for pp and 40--70\% for p--Pb. 
A good agreement between the two correlation functions is observed. This reveals the genuine behavior of correlations, as the rescaled two-particle cumulant eliminates the 1/N scaling.  Although the comparison between the two systems is not a perfect match, since the values of ${\rm d}N_{ch}/{\rm d}\eta$ are not perfectly equal, they agree within the experimental uncertainties. 
\section{Conclusion}
This analysis concerns the two-particle angular correlation in $\Delta y,\Delta\varphi$ space for pp and p--Pb collisions at the LHC energies using alternative definitions, i.e., the probability ratio and the rescaled two-particle cumulant. The results reveal that the correlation functions obtained using the latter, increase with the multiplicity. Furthermore, the implementation of the following definition allows us to explore in depth the physical phenomena that contribute to the overall shape of the correlation functions by eliminating the trivial scaling of 1/N. When comparing the results between the two collision systems, a multiplicity dependence is observed where the number of correlated particles grows very fast compared with those of the underlying event. The anticorrelation between proton pairs is still present and becomes stronger at high multiplicities. The comparison of the correlation function with the rescaled two-particle cumulant definition in the pp and p--Pb systems shows good agreement within the experimental uncertainties for the closest multiplicity classes.

\section{Acknwledgments}
This work was supported by the Polish National Science Centre under agreements \textcolor{blue}{UMO-2021/43/D/ST2/02214} and \textcolor{blue}{UMO-2022/45/B/ST2/02029}, by the Polish Ministry for Education and Science under agreements no. \textcolor{blue}{2022/WK/01} and \textcolor{blue}{5236/CERN/2022/0}, as well as by the \textcolor{blue}{IDUB-POB-FWEiTE-3} project granted by Warsaw University of Technology under the program Excellence Initiative: Research University (ID-UB).
\renewcommand{\refname}{} 
\renewcommand\refname{}


\section{References}
\vspace{-1cm}
\begin{thebibliography}{0}
\bibitem{Adam_2017} \BY{ALICE Collaboration}
  \IN{The European Physical Journal C}{77}{2017}{569};

\bibitem{starcoll} \BY{STAR Collaboration}
  \IN{Physical Review C}{101}{2020}{014916};
\bibitem{DRuggiano} \BY{Ruggiano D.}
  \IN{arXiv}{2023}{2311.09833};

\bibitem{probability} \BY{Kisiel A.}
  \IN{Phys. Rev. C}{81}{2010}{064906};

\bibitem{rescaled} \BY{STAR Collaboration}
  \IN{Phys. Rev. C}{86}{2012}{064902};
  
\bibitem{detectorsdescription} \BY{ALICE Collaboration}
  \IN{Journal of Instrumentation}{3}{2008}{S08002};

\bibitem{MBiasTrigger} \BY{Krivda M. et al.}
  \IN{Journal of Instrumentation}{7}{2012}{C01057};

\bibitem{AnnichilationSTAR} \BY{STAR Collaboration}
  \IN{Phys. Rev. C}{101}{2020}{014916};

\bibitem{Unfolding} \BY{Graczykowski, \L{}. and Janik, M.}
   \IN{Phys. Rev. C}{104}{2021}{054909};
   
\bibitem{dNchdetavaluespp} \BY{ALICE Collaboration}
  \IN{The European Physical Journal C}{80}{2020}{1434-6052};

\bibitem{dNchdetavaluespPb} \BY{ALICE Collaboration}
  \IN{Physics Letters B}{728}{2015}{0370-2693};

\end{thebibliography}
\end{document}